\documentclass[fleqn,twoside]{article}
\usepackage{espcrc2}

\usepackage{graphicx,amssymb}
\usepackage[figuresright]{rotating}

\newcommand{\AmS}{{\protect\the\textfont2
  A\kern-.1667em\lower.5ex\hbox{M}\kern-.125emS}}

\title{Finite temperature phase transition, adjoint Polyakov loop and topology in $SU(2)$ LGT\thanks{Talk given by A. Barresi at Lattice2001, Berlin. HU-EP-01/41}}

\author{Andrea Barresi, Giuseppe Burgio\thanks{Address from Nov. $1^{\rm st}$ 2001: School of Mathematics, Trinity College, Dublin 2, Ireland.}, Michael M\"uller-Preussker\\
$\;$\\
Humboldt-Universit\"at zu Berlin, Institut f\"ur Physik, 10115, Germany}
       
\begin{document}

\begin{abstract}
We investigate the phase structure of pure $SU(2)$ LGT at finite temperature in the mixed fundamental and adjoint representation modified with a $\mathbb{Z}_2$ 
monopole chemical potential. The decoupling of the finite temperature phase transition from unphysical zero temperature bulk phase transitions is analyzed with special emphasis on the continuum limit. The possible relation of the adjoint Polyakov loop to an order parameter for the finite temperature phase transition and to the topological structure of the theory is discussed.
\vspace{1pc}
\end{abstract}

\maketitle

\section{INTRODUCTION}

Pure $SU(N)$ lattice gauge theories within the fundamental representation of 
the gauge group show a finite temperature deconfinement phase transition together with the breaking of a global $\mathbb{Z}_N$ center symmetry. But if confinement is a feature of the Yang-Mills continuum degrees of freedom it should be independent of the group representation for the lattice action. As Polyakov's center symmetry breaking mechanism is available only to half-integer representations of the group, a finite temperature investigation of Wilson's action for $SU(2)$ in the adjoint representation, i.e. $SO(3)$, might offer interesting insight to the present understanding of confinement. 

The $SU(2)$ mixed fundamental-adjoint action was originally studied by  Bhanot and Creutz \cite{1BC81}:
\begin{equation}
\label{eq1}
S\!=\!\sum_{P}\Bigg[\!\beta_{A}\Bigg(1-\frac{\mathrm{Tr}_{A}U_{P}}{3}\Bigg)+\beta_{F}\Bigg(1-\frac{\mathrm{Tr}_{F}U_{P}}{2}\Bigg)\!\Bigg]
\end{equation}
They found the well known non-trivial phase diagram characterized by
 first order $T=0$ bulk phase transition lines. A similar phase diagram is shared by $SU(N)$ theories with $N\ge 3$ \cite{2BC81}. 

Halliday and Schwimmer \cite{1HS81} found a similar phase diagram using a Villain discretization for the center blind part of action (\ref{eq1})
\begin{equation}
S=\!\!\sum_{P}\!\!\Bigg[\!\beta_{V}\Bigg(\!1-\frac{\sigma_{P}\mathrm{Tr}_{F}U_{P}}{2}\!\Bigg)\!+\beta_{F}\Bigg(\!1-\frac{\mathrm{Tr}_{F}U_{P}}{2}\!\Bigg)\!\Bigg]
\end{equation}
$\sigma_{P}$ being an auxiliary $\mathbb{Z}_2$ plaquette variable. By defining $\mathbb{Z}_2$ magnetic monopole and electric vortex densities 
$M=1-\langle\frac{1}{N_{c}}\sum_{c}\sigma_{c}\rangle$,
$E=1-\langle\frac{1}{N_{l}}\sum_{l}\sigma_{l}\rangle$ with
$\sigma_{c}=\prod_{P\epsilon\partial c}\sigma_{P}$ and
$\sigma_{l}=\prod_{P\epsilon\hat{\partial} l}\sigma_{P}$
they argued that the bulk phase transitions were caused by condensation of these lattice artifacts. They also suggested \cite{2HS81} a possible suppression mechanism via the introduction of chemical potentials of the form
$\lambda\sum_{c}(1-\sigma_{c})$ and $\gamma\sum_{l}(1-\sigma_{l})$.

Recently Gavai and Datta \cite{1G99} explicitely realized this suggestion, studying the 
$\beta_{V}-\beta_{F}$ phase diagram as a function of $\lambda$ and $\gamma$. They found
lines of second order finite temperature phase transitions crossing the 
$\beta_V$ and $\beta_F$ axes for $\lambda\ge 1$ and $\gamma\ge 5 $.
In the limiting case $\beta_{F}=0$ and $\gamma=0$, i.e. $SO(3)$ theory with a $\mathbb{Z}_2$ monopole chemical potential, a quantitative study is difficult because of the lack of an order parameter. The $\mathbb{Z}_2$ global symmetry remains trivially unbroken. A thermodynamical approach \cite{2G99} shows a steep rise in the energy density for asymmetric lattices with $N_{\tau}=2,4$ and a peak in the specific heat at least for $N_{\tau}=2$, supporting the idea of a second order deconfinement phase transition. The authors have seen the adjoint Polyakov loop 
to fluctuate around zero below the phase transition and to take the values $1$ and $-\frac{1}{3}$ above the phase transition as
$\beta_V\to \infty$.    

\section{ADJOINT ACTION WITH CHEMICAL POTENTIAL}

We study an adjoint representation Wilson action modified by a chemical potential suppressing the $\mathbb{Z}_2$ magnetic monopoles
\begin{equation}
S=\frac{4}{3}\beta_{A}\sum_{P}\Bigg(1-\frac{\mathrm{Tr}_{F}^{2}U_{P}}{4}\Bigg)+\lambda\sum_{c}(1-\sigma_{c})
\end{equation}

The link variables are taken in the fundamental representation only to improve the speed of our simulations, after checking that with links represented by $SO(3)$ matrices nothing changes. A standard Metropolis algorithm is used to update the links.
The term $\sigma_{c}=\prod_{P\epsilon\partial c}\mathrm{sign}(\mathrm{Tr}_{F}U_{P})$ is completely center blind, i.e. $U_{\mu}(x)\rightarrow -U_{\mu}(x)\Rightarrow\sigma_{c}\rightarrow\sigma_{c}\;\;\;\forall \mu,x,c$. 
\begin{figure}[htb]
\label{fig:btlam}
\includegraphics[width=0.45\textwidth]{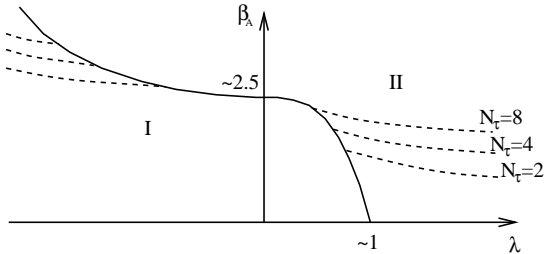}
\caption{The phase diagram in the $\beta_A-\lambda$ plane for various $N_\tau$.}
\end{figure}
Fig. 1 shows the phase diagram in the $\beta_A-\lambda$ plane at finite temperature. The two phases (I-II) are separated by a bulk first order line at which $\mathbb{Z}_2$ monopoles condense, phase I being continously connected with the physical $SU(2)$ phase as $\beta_F$ is turned on. Finite temperature lines, at which $\langle L_A \rangle$  shows a jump, cross the plane more or less horizontally. Putting aside the order parameter problem, the scaling behaviour at the critical temperature $T_c\equiv\frac{1}{aN_\tau}$ as a function of $\beta_A$ and $\lambda$ turns out difficult in phase I, whereas in phase II it shows a nice scaling behaviour in $\beta_A$ at fixed $\lambda \gtrsim 1$.

\section{SYMMETRY AND ORDER PARA\-METER}

A quantitative study of the observed finite temperature transition is viable either relying on pure thermodynamical quantities \cite{2G99} or defining a reasonable order parameter, i.e. by understanding the underlying symmetry breaking mechanism, if any. The only hints we have are the change in the distribution of the adjoint Polyakov line operator $\frac{1}{3}\mathrm{Tr} L_A(\vec{x})$ and the values it takes in the continuum limit. After maximal abelian gauge (MAG) \cite{MAG} and abelian projection it is indeed possible to establish an exact global symmetry which can be broken at the phase transition and a related order parameter.
Taking 
$$O_{\mu}(x)=I+ \sin 2 \theta_{\mu}(x) T_3 + (1-\cos 2 \theta_{\mu}(x))T_3^2$$
as the projected link in the adjoint theory, with $\vec{T}$ the adjoint representation generators of the Lie algebra, it is easy to see that the ``parity'' operator $P=I+2 T_3^2$ acting on all links living at fixed time as 
$$P O_{\mu}(x)=I- \sin 2 \theta_{\mu}(x)T_3 + (1+\cos 2 \theta_{\mu}(x))T_3^2$$
leaves all the plaquettes (and thus the action) invariant, while changing the Polyakov line. 
If $\Theta_L  (\vec{x})  =\sum_{n=0}^{N_{\tau}-1} \theta_4(\vec{x}+\,n\,a\,\hat{4})$ is the Polyakov line global abelian phase, then for the spatial average $\langle \mathrm{Tr}L_A\rangle=1+2\langle \mathrm{cos} 2 \Theta_L  (\vec{x})\rangle$ and $\langle \mathrm{Tr}PL_A\rangle=1-2\langle \mathrm{cos} 2 \Theta_L  (\vec{x})\rangle$. If this symmetry is broken at the phase transition, then $\langle \mathrm{Tr}L_A\rangle=1$ below and $\langle \mathrm{Tr}L_A\rangle=1\pm2\Delta$ above, with $\Delta=\langle \mathrm{cos} 2 \Theta_L\rangle$. Thus, a reasonable order parameter should be $|\Delta|=\frac{1}{2}|\langle\mathrm{Tr}L_A\rangle-1|$.
\begin{figure}[htb]
\label{fig:dist}
\begin{center}
\includegraphics[totalheight=1.2in]{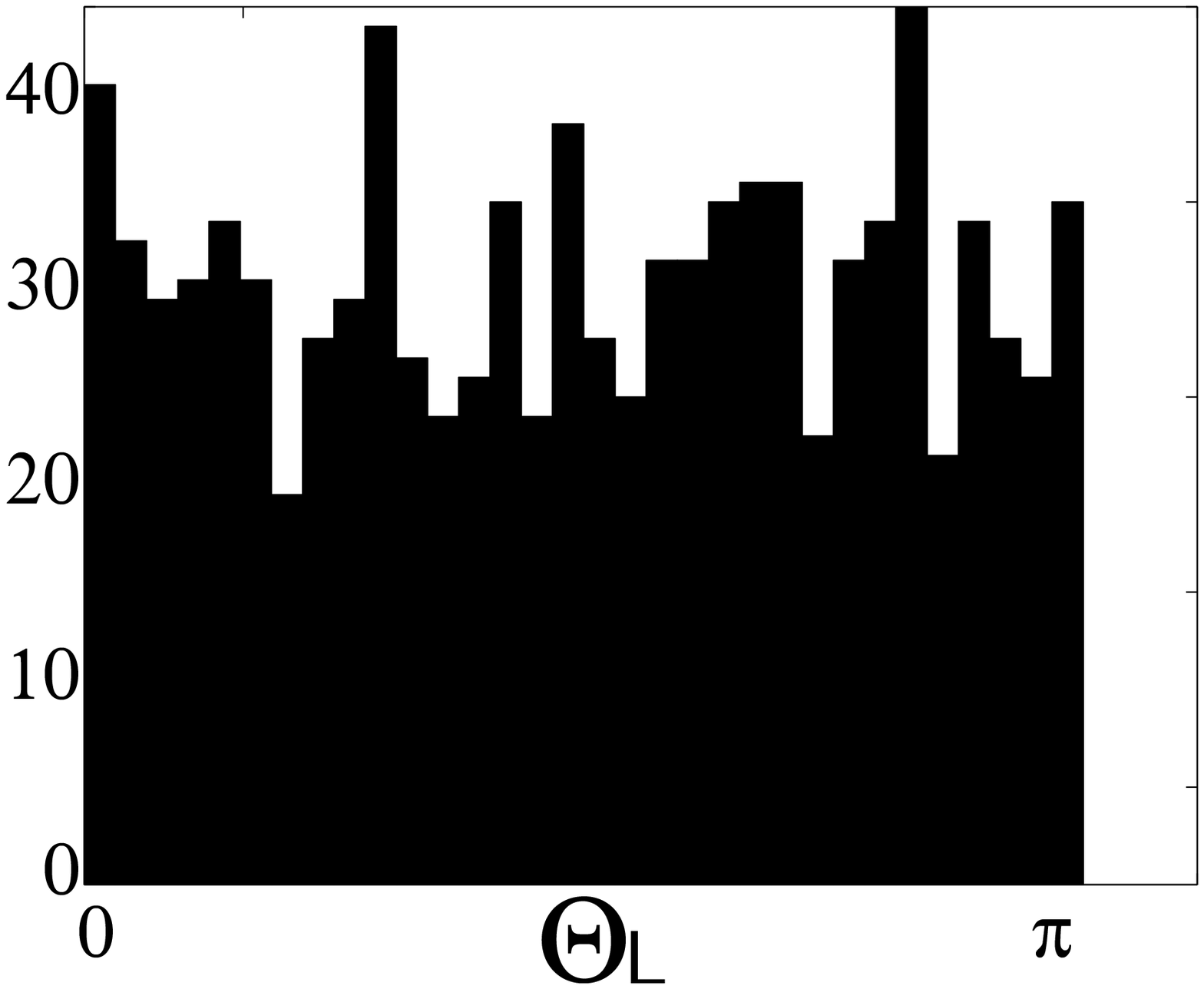}
\end{center}
\includegraphics[totalheight=1.2in]{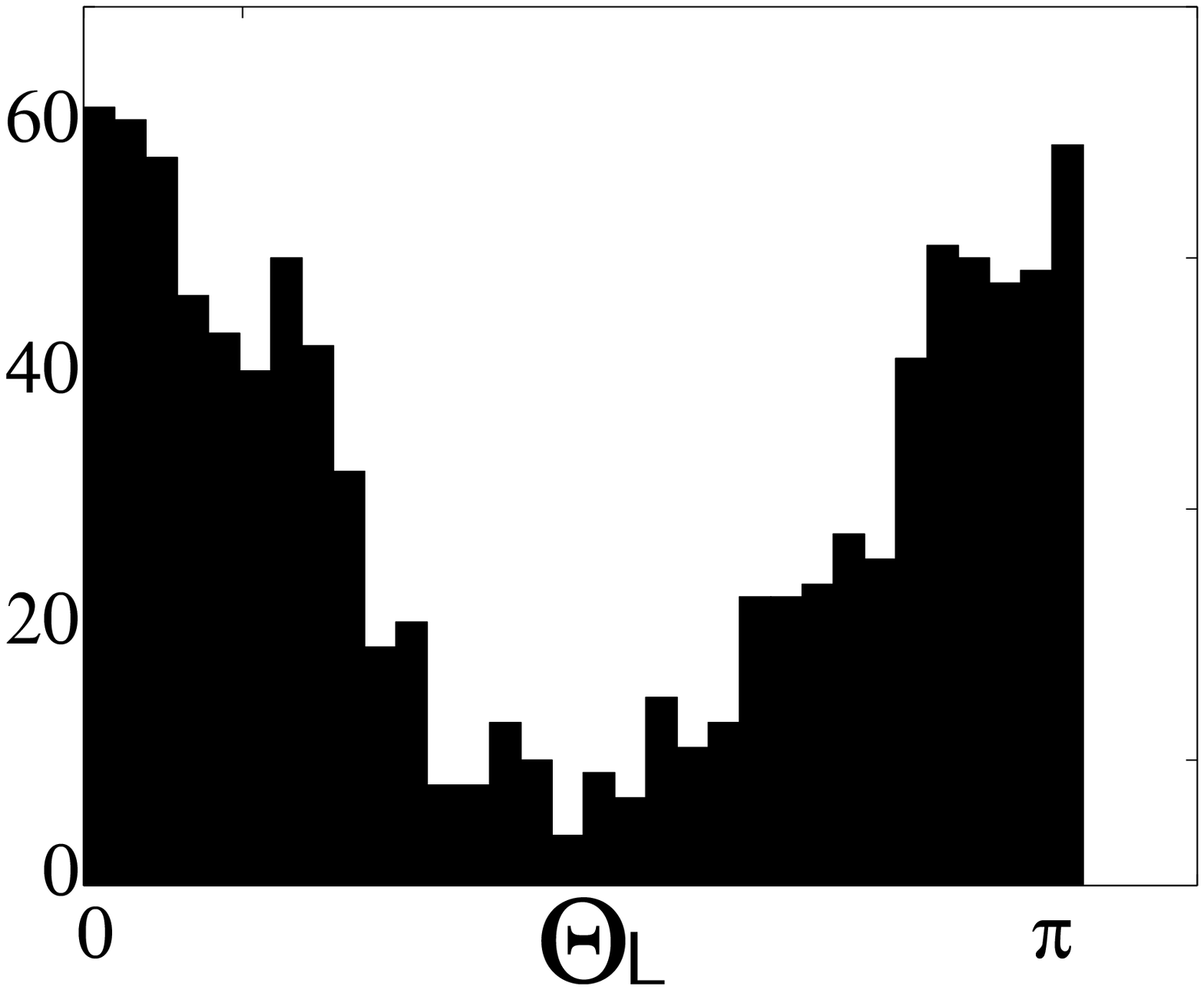}
\includegraphics[totalheight=1.2in]{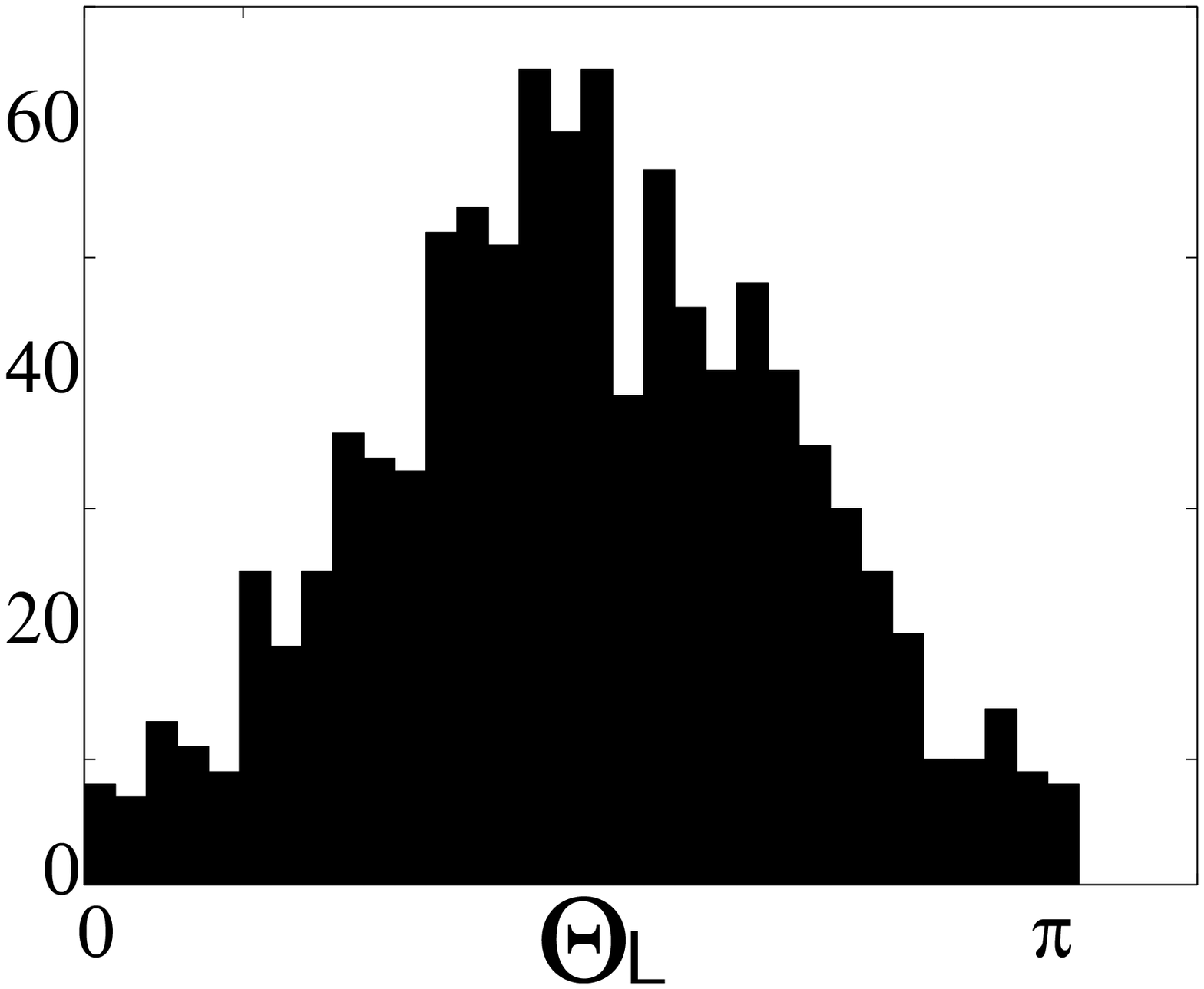}
\caption{$\Theta_L(\vec{x})$ volume distribution below ($\Delta=0$) and above the transition ($\Delta=\pm 1$) for typical configurations.}
\end{figure}

Fig. 2 shows the volume distribution of the Polyakov line angle at the phase transition for some typical configurations. Although such a sharp change can be observed also for the full ${\rm Tr} \mathrm{L}_A(\vec{x})$ distribution, in the latter case a quantitative analysis is made difficult by the asymmetry of the values at which it peaks. In the abelian projected case, after MAG, $\Theta_L(\vec{x})$ is clearly flat below the phase transition, peaking around $0 (\pi)$ and $\frac{\pi}{2}$ above. In Fig. 3 the proposed order parameter is plotted as a function of $\beta_A$ for $\lambda=1$ and $N_\tau=4$. A singular behaviour around $\beta_A\simeq 1$ is starting to show at $V=16^3$. At $N_\tau=6$ the critical $\beta_A$ increases by roughly $35\%$. 
\begin{figure}[htb]
\label{fig:magn}
\includegraphics[totalheight=2.02in]{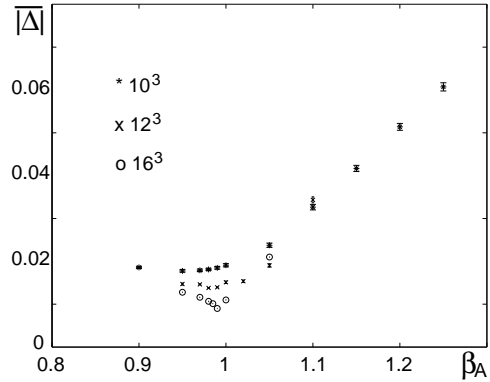}
\caption{Ensemble average of $|\Delta|$ vs. $\beta_A$.}
\end{figure}

The results show that the proposed symmetry breaking mechanism is plausible and that the order parameter behaves as one expects for a $2^{\rm nd}$ order transition, although more data at higher volumes and a study of the susceptibility would be necessary to asses such statements. The analy\-sis of Binder cumulants is also feasible with our definitions. A study of the critical exponents and of the cluster properties of $\langle \mathrm{Tr}L_A \rangle$ would be as well interesting in order to establish whether the features of such a system are similar to those of the usual $SU(2)$ phase transition. All these questions will be addressed in a forthcoming paper.

\section{CONCLUSIONS}

We have studied the phase diagram of the adjoint Wilson action with a chemical potential $\lambda$ for the $\mathbb{Z}_2$ magnetic monopoles. The finite temperature phase transition can be decoupled from the bulk phase transition both for positive and negative $\lambda$. The scaling behaviour of $\beta_A$ with $N_\tau$ is established in both cases, although the type I phase presents some difficulties in taking the continuum limit. In the context of abelian dominance we propose a symmetry breaking mechanism and an order parameter for the phase transition, giving promising results for numerical simulations. A deeper numerical analysis and possible extensions of the definitions will be the subject of a forthcoming paper. This work was funded by a EU-TMR network under the contract FMRX-CT97-0122 and by the DFG-GK 271.

\end{document}